\documentclass[10pt,twocolumn,aps,pre,superscriptaddress]{revtex4} %

\usepackage{amsmath}
\usepackage{graphicx}
\usepackage{color}
\usepackage{upgreek}
\usepackage[linkcolor = blue, citecolor = blue, urlcolor = blue, colorlinks = true]{hyperref}
\usepackage{amssymb}

\usepackage{bm}

\begin{document}

\author{Joost de Graaf}
\email{jgraaf@icp.uni-stuttgart.de}
\affiliation{SUPA, School of Physics and Astronomy, The University of Edinburgh, King's Buildings, Peter Guthrie Tait Road, Edinburgh, EH9 3FD, United Kingdom}

\author{Joakim Stenhammar}
\email{joakim.stenhammar@fkem1.lu.se}
\affiliation{Division of Physical Chemistry, Lund University, P.O. Box 124, S-221 00 Lund, Sweden}

\title{Lattice-Boltzmann Simulations of Microswimmer-Tracer Interactions}

\date{\today}

\begin{abstract}
Hydrodynamic interactions in systems comprised of self-propelled particles, such as swimming microorganisms, and passive tracers have a significant impact on the tracer dynamics compared to the equivalent ``dry'' sample. However, such interactions are often difficult to take into account in simulations due to their computational cost. Here, we perform a systematic investigation of swimmer-tracer interaction using an efficient force/counter-force based lattice-Boltzmann (LB) algorithm [J. de Graaf~\textit{et al.}, J. Chem. Phys.~\textbf{144}, 134106 (2016)] in order to validate its ability to capture the relevant low-Reynolds-number physics. We show that the LB algorithm reproduces far-field theoretical results well, both in a system with periodic boundary conditions and in a spherical cavity with no-slip walls, for which we derive expressions here. The force-lattice coupling of the LB algorithm leads to a ``smearing out'' of the flow field, which strongly perturbs the tracer trajectories at close swimmer-tracer separations, and we analyze how this effect can be accurately captured using a simple renormalized hydrodynamic theory. Finally, we show that care must be taken when using LB algorithms to simulate systems of self-propelled particles, since its finite momentum transport time can lead to significant deviations from theoretical predictions based on Stokes flow. These insights should prove relevant to the future study of large-scale microswimmer suspensions using these methods.
\end{abstract}

\maketitle

\section{\label{sec:intro}Introduction}

Suspensions of biological ``microswimmers'', usually consisting of swimming bacteria, algae, and protozoa, exhibit many interesting properties, both from a biological and from a basic statistical physics point of view~\cite{lauga09,Marchetti-2013,Elgeti-2015,Bechinger-2016}. One set of problems that has attracted particular interest over the last decade is the enhanced diffusion of non-swimming (``passive'') tracer particles suspended in a bacterial or algal bath~\cite{Wu-2000, Underhill-2008, Leptos-2009, Dunkel-2010, Thiffeault-2010, Pedley-2010, Lin-2011, Valeriani-2011, Pushkin13mix, Pushkin-2013, Jepson-2013, Mino-2013,  Morozov-2014, Thiffeault-2015, Krishnamurthy-2015, Jeanneret-2016, Mueller16}, compared to that expected from thermal fluctuations alone. This phenomenon has been extensively analyzed theoretically and rationalized in terms of characteristic hydrodynamic scattering events between the tracer and the swimmer flow-field~\cite{Pushkin13mix}. 

While the flow field close to a microswimmer is complex and specific to each organism~\cite{Drescher10, Drescher11}, the hydrodynamic far-field flow can readily be described using a superposition of fundamental solutions to the incompressible Stokes equation for the fluid velocity $\boldsymbol{u}(\boldsymbol{r})$~\cite{Spagnolie-2012}:
\begin{align}
\label{eq:Stokes} \mu \underline{\Delta} \boldsymbol{u}(\boldsymbol{r}) - \boldsymbol{\nabla} p(\boldsymbol{r}) &= - \boldsymbol{F}(\boldsymbol{r}) ;\\
\label{eq:Stokclos} \boldsymbol{\nabla} \cdot \boldsymbol{u}(\boldsymbol{r}) &= 0.
\end{align}
Here, $\boldsymbol{r}$ is the position, $p(\boldsymbol{r})$ is the pressure, $\mu$ is the dynamic viscosity, $\underline{\Delta}$ is the vector Laplacian, and $\boldsymbol{F}$ is a volume force distribution acting on the fluid. These equations neglect any time-dependence of the flow, by discarding the inertial terms present in the full Navier-Stokes equation. This overdamped approximation is highly accurate for treating organisms swimming at the microscale, since virtually all such swimmers operate in the regime of negligible Reynolds numbers, as defined by
\begin{align}
\label{eq:Re} \mathrm{Re} &= \frac{\rho v_{s} \ell }{\mu},
\end{align} 
with $\rho$ the mass density of the fluid, $v_{s}$ the swimming velocity, and $\ell$ is a relevant length scale of the problem. For bacteria and algae, the Reynolds number of an isolated swimmer is usually of the order $\mathrm{Re} = 10^{-5}-10^{-2}$~\cite{lauga09}, where we take $\ell$ to be the length of the organism. This means that friction completely dominates inertia and that the flow field throughout the system can be assumed to respond instantaneously to changes in the boundary conditions. 

Since microswimmers are force-free --- provided gravitational forces are neglected --- the leading-order hydrodynamic singularity of such a swimmer is typically that of a point hydrodynamic force dipole (or, equivalently, symmetric Stokes doublet or stresslet):
\begin{align}
\label{eq:bulkStresslet} \boldsymbol{u}(\boldsymbol{r}) &= \frac{\kappa}{8\pi \mu r^{2}}\left( 3(\hat{\boldsymbol{p}}\cdot \hat{\boldsymbol{r}})^{2} - 1 \right)\hat{\boldsymbol{r}},
\end{align}
where $\kappa$ is the stresslet strength, $\hat{\boldsymbol{p}}$ is the swimmer orientation, and $\hat{\boldsymbol{r}}$ gives the separation unit vector between the swimmer and the observation point $\boldsymbol{r}$. By construction, positive values of $\kappa$ correspond to rear-actuated microswimmers (``pushers'', extensile) such as \textit{E. coli}~\cite{Drescher11}, and $\kappa < 0$ represents front-actuated organisms (``pullers'', contractile) such as \textit{Chlamydomonas}~\cite{Drescher10}. Since a real microswimmer will have a finite separation between the force points, the description of a microswimmer flow field as that of a point stresslet is only valid at distances appreciably larger than the typical size of the swimmer. Nevertheless, this minimal stresslet-based model has proven accurate in numerically describing collective phenomena in microswimmer suspensions~\cite{Dunkel-2010, Pushkin-2013, Morozov-2014}, while still being simple enough to provide some analytical tractability~\cite{Menzel-2016, Saintillan-2008, Subramanian-2009}.

Computationally, hydrodynamic aspects of microswimmer suspensions have been studied using a variety of fluid-dynamical solvers, including Stokesian dynamics~\cite{Ishikawa-2010, Evans-2011}, multi-particle collision dynamics (MPCD)~\cite{Goetze-2010, zoettl14}, boundary-element methods~\cite{Ishimoto-2013, Li-2014}, and lattice-Boltzmann (LB) simulations~\cite{nash10, Pagonabarraga-2013}. Treating Stokes flows has great advantages from a theoretical point of view, but is often difficult to achieve in simulations. Methods such as LB and MPCD are constructed to solve the full Navier-Stokes equation, including the inertial term. This can lead to difficulties when treating microswimmers using such numerical fluid dynamics solvers, especially in comparing to theoretical results, as we will examine in detail in this manuscript for the case of LB. Collective motion, such as bacterial turbulence~\cite{Saintillan-2012, Dunkel-2013, Krishnamurthy-2015, Secchi-2016}, is particularly strongly impacted by these limitations, because the relevant length-scale $\ell$ in Eq.~\eqref{eq:Re} should then be the typical \textit{vortex size}, which can be 1-2 orders of magnitude larger than the size of the individual swimmer. This can push the relevant Reynolds number outside the Stokes flow regime ($\mathrm{Re} < 0.1$) for typical LB parameters, while in the physical system the Reynolds number of the vortex motion remains negligible. The reason is that the Reynolds number of the swimmers is typically taken considerably larger in LB studies than in the experiment in order to speed up the simulations. Thus, in the simulations care needs to be taken in order to keep all relevant Reynolds numbers small. 

\begin{figure}[!htb]
\centering 
  \includegraphics[scale=1.0]{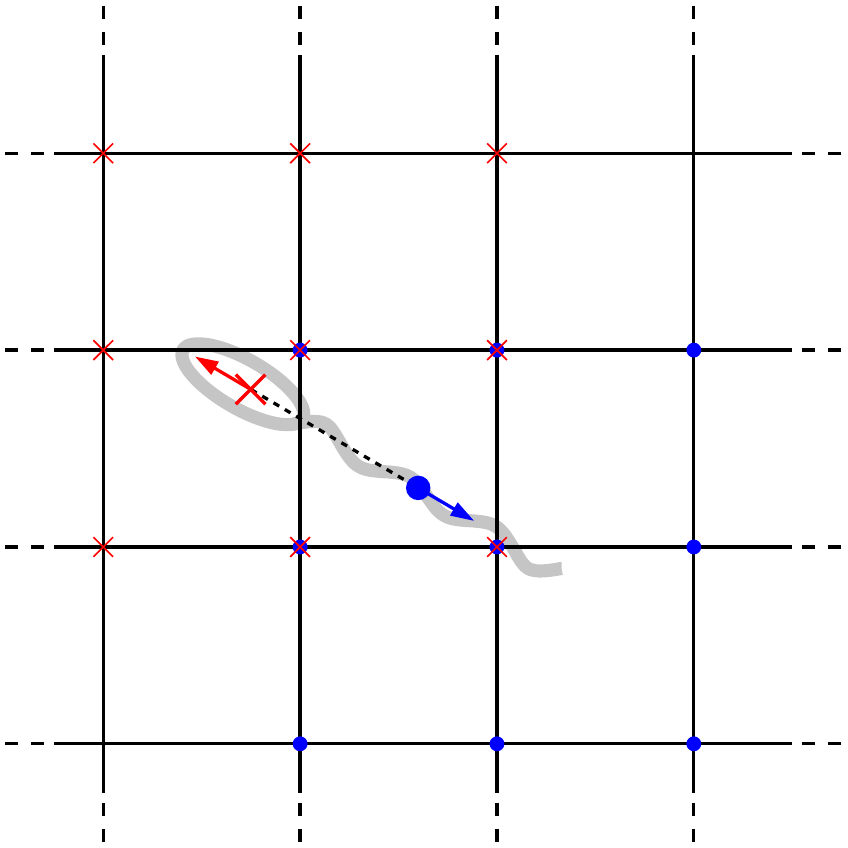} 
  \caption{\label{fig:model}Two-dimensional (2D) schematic representation of a force/counter-force swimmer that models a pusher microorganism such as \textit{E. coli}. The thin black lines indicate a part of the LB grid. The swimmer is represented by two forces (indicated by a red and a blue arrow), and moves in the direction of the red arrow. The off-lattice forces are interpolated onto the lattice using a 3-point scheme; the interpolated forces are denoted with smaller dots and crosses of the same color as the off-lattice force. Similarly, fluid velocities are interpolated from the lattice onto the position of the swimmer. Since the typical separation between our force and counter-force point is one lattice spacing (dotted line), the interpolation points overlap, albeit with different weights from the two off-lattice points.}
\end{figure} 

In this study, we will employ an LB force coupling method put forward in Ref.~\cite{degraaf16a}, together with a simple force/counter-force microswimmer description, which is a well-established minimal model of microswimmers~\cite{Hernandez-Ortiz05, saintillan07, swan11, Lushi014, singh15, Menzel-2016}. Each swimmer is described by two equal and opposite forces $\pm \boldsymbol{f}$ separated by a finite distance $l$, as illustrated in Fig.~\ref{fig:model}, such that the flow field of each swimmer exactly reduces to that of Eq.~\eqref{eq:bulkStresslet} in the limit $l \rightarrow 0$, when $fl \equiv \kappa$ is kept constant. We will systematically evaluate the LB scheme with respect to the hydrodynamic interaction between a single swimmer and a tracer, a problem which is central to the description of enhanced diffusion in microswimmer suspensions. We do so by comparing to the corresponding exact solutions of the Stokes equation for the same problem, as established in Ref.~\cite{Dunkel-2010}. 

First, we consider the near-field flows,~\textit{i.e.}, tracer trajectories for short swimmer-tracer separations. For this situation, the LB force-fluid coupling leads to a short-ranged regularization of the stresslet flow fields, when compared to the exact (singular) result. We show that the inherent regularization present in the LB method can be well-matched to a simple theoretical regularization of the stresslet.

Second, in the far field, we find excellent agreement between LB results and the theoretical predictions for a system with periodic boundary conditions (PBCs)~\cite{degraaf16c} and for a finite, spherical cavity with no-slip walls --- we derive expressions for the latter in Appendix \ref{sec:cavity}. We note that the influence of the boundary conditions is strikingly large, even for swimmer-tracer separations significantly smaller than the system dimensions. Surprisingly, there are more similarities between the system with PBCs and the finite-sized cavity, than there are between these two and the infinite bulk system. 

Finally, we evaluate the effect of momentum retardation due to non-zero Reynolds numbers. We find that retardation of the hydrodynamic interactions strongly perturbs the tracer trajectory for $\mathrm{Re} > 0.1$, with $\ell$ appropriately chosen to represent the length scale relevant to the problem.

\section{\label{sec:methods}Model and Methods}

\begin{figure*}[!htb]
\centering 
  \includegraphics[scale=1.0]{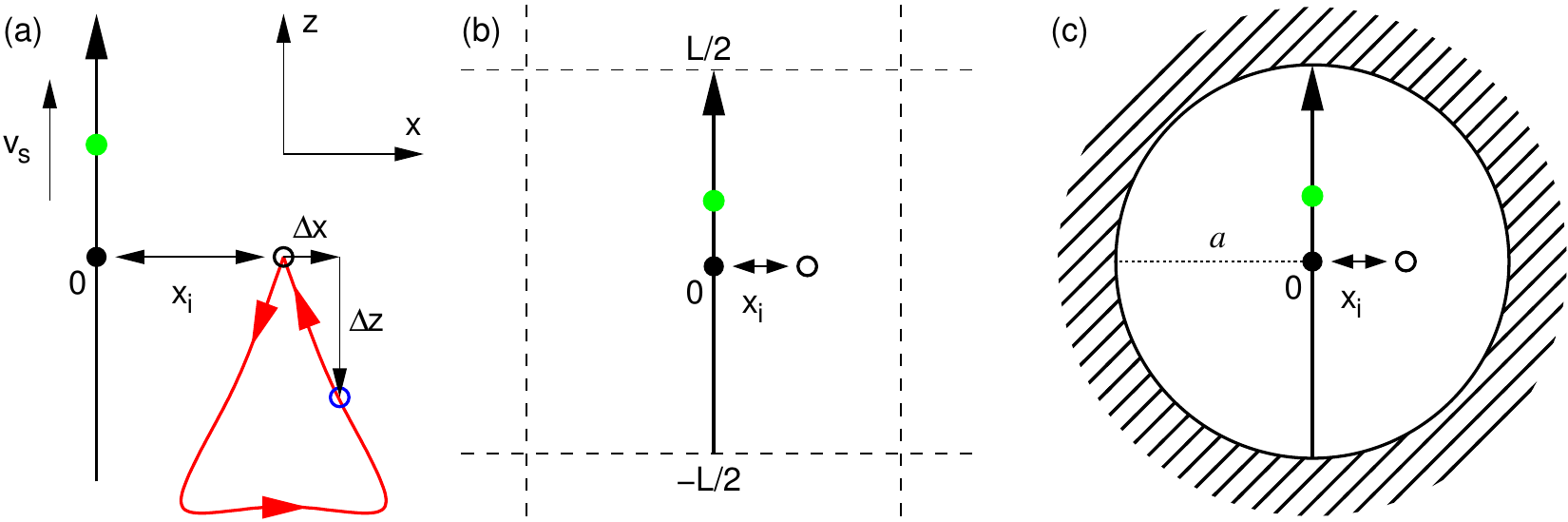} 
  \caption{\label{fig:geom}The various geometries in which we study swimmer-tracer scattering. (a) The swimmer (green circle) moves along the $z$-axis with constant speed $v_{s}$. The tracer (blue circle) is initially located on the $x$-axis at position $x_{i}$ and its trajectory due to the flow field produced by the swimmer is parameterized by $\Delta x$ and $\Delta z$. The red line with arrows shows a sketch of a typical trajectory of a tracer being advected in the flow field of a puller-type swimmer. (b) A 2D representation of a cubic system with edge length $L$ with PBCs (dashed lines). (c) 2D representation of a spherical cavity of radius $a$ with no-slip walls (patterned).}
\end{figure*} 

The behavior of swimmers and tracers is governed by only a few quantities, which are shown in Fig.~\ref{fig:geom}a. In all of our studies, the swimmer moves along the $\boldsymbol{\hat{z}}$-direction with constant swimming speed $v_{s}$~\footnote{In our work we also simulate swimmers near solid walls. The presence of such a wall influences the swimming speed, as described by Fax{\'e}n's laws. However, we found the effect to be minimal for the trajectories that we considered and a constant swimming speed is therefore a good approximation.}, and the tracer is initially located at $\boldsymbol{r}_{i} = x_{i} \boldsymbol{\hat{x}}$. Due to the flow field induced by the swimmer, the tracer moves along a trajectory parametrized by $\Delta x$ and $\Delta z$. In order to make a comparison between various forms of boundary conditions on the advection of tracers, we consider two different simulation geometries. The first is a cubic box with edge length $L$ and PBCs in all three directions, see Fig.~\ref{fig:geom}b. The second is a spherical cavity of radius of $a$ with no-slip (zero velocity) walls, see Fig.~\ref{fig:geom}c. 

\subsection{\label{sub:lbmodel}LB Simulations}

For the LB simulations, a graphics processing unit (GPU) implementation~\cite{roehm12} was used. We employ a fluctuating multiple relaxation time (MRT) collision operator~\cite{dhumieres02}, although here we only consider quiescent (unthermalized) fluids. All simulations were performed using the MD software \textsf{ESPResSo}~\cite{limbach06a,arnold13a}, using a fluid density of $\rho = 1.0$, lattice spacing $\Delta L = 1.0$, time step $\Delta t = 0.005$, kinematic viscosity $\nu = 1.0$, and a bare particle-fluid friction of $\zeta_{0} = 25$ --- we use LB units throughout. We refer the reader to Ref.~\cite{fischer15} for a detailed description of the dimensionless numbers that specify the fluid properties to which these choices correspond. The LB parameters used here are identical to those used in Refs.~\cite{fischer15, degraaf15b, degraaf16a, degraaf16b} and can therefore be expected to faithfully reproduce hydrodynamics in a variety of geometries. 

We employ the approach discussed in Ref.~\cite{degraaf16a} to model the hydrodynamic interactions between swimmers and tracers. In this approach, the microswimmer's ``body'' consists of a single point particle with an applied point force that couples to the LB fluid via the scheme due to Ahlrichs and D{\"u}nweg~\cite{ahlrichs99}. In order to make the system force free, as is the case for self-propelled objects, we apply a counter force to the fluid, separated from the body a distance $l$ away, where $l$ is comparable to the lattice spacing, also see Ref.~\cite{degraaf16a}. The direction of the forces and the position of the counter-force point co-rotate with the swimmer, thus representing the flagella and the microswimmer body, respectively, see Fig.~\ref{fig:model}. The friction resulting from the LB coupling between the body and the fluid results in a fixed swimming speed $v_{s}$. The forces and swimmer velocities are interpolated between the swimmers and the lattice using a 3-point stencil~\cite{ladd94}, which has been shown to significantly reduce lattice artifacts compared to the traditional 2-point one~\cite{degraaf16a}. The passive tracer particles are modeled using the same LB coupling,~\textit{i.e.}, a single bead that does not experience an external force. Through the coupling, this implies that the bead is simply advected by the fluid flow. 

Unless otherwise specified, we apply a force of $f = 0.01$ to the swimmer and a counter-force of equal magnitude at a distance of $l = 1$ away from the swimmer body. This causes the swimmer to move with a speed of $v_{s} \approx 5.4 \times 10^{-4}$ and gives rise to a hydrodynamic dipole moment (stresslet) of magnitude $\kappa = f l = 0.01$. An independent measurement of the dipole strength by Legendre-Fourier decomposition of the swimmer's flow field, see Ref.~\cite{degraaf16a}, yielded $\kappa \approx 1.4 \times 10^{-2}$, which is an acceptable deviation from $\kappa = f l$, given the fairly large uncertainty ($\gtrsim 20\%$) connected with this measurement. The associated single-swimmer Reynolds number is $\mathrm{Re} = 5.4 \times 10^{-4}$.

For simulations employing PBCs, a cubic box with side length $L = 100$ was employed throughout, while for spherical cavity simulations, a cavity radius of $a = 50$ was used. The latter geometry was implemented using a zero-velocity boundary condition based on the bounce-back algorithm~\cite{frisch86}, emulating the effect of no-slip walls. Since we employ a 3-point interpolation stencil for the forces and velocities, swimmer trajectories were started at $-(a - 2)\boldsymbol{\hat{z}}$ and terminated when the swimmer reached $(a - 2)\boldsymbol{\hat{z}}$, thus preventing undesirable wall-swimmer interactions~\cite{degraaf15b}.

We finally note that our model is similar to the microswimmer model of Nash~\textit{et al.}~\cite{nash08, nash10}. The main exception is that the latter method instead \textit{imposes} a swimming speed $v_{s}$ through the Stokes friction for a sphere with a predefined radius, thus advancing the particles through overdamped dynamics. Due to the similarities of the two methods, we however expect that the results obtained here should also be applicable to that force-coupling scheme. 

\subsection{\label{sub:compare}Comparisons with Theory}

We compare the results of our simulations to theoretical predictions obtained by explicitly solving the Stokes equation in various geometries and using different approximations. In all cases the theoretical tracer trajectory is determined by numerically solving the coupled differential equations
\begin{align}
\label{eq:tracer}  \dot{\boldsymbol{r}}_{\mathrm{tr}}(t) &= \boldsymbol{u}(\boldsymbol{r}_{\mathrm{tr}}(t) - \boldsymbol{r}_{\mathrm{s}}(t)); \\
\label{eq:swimmer} \dot{\boldsymbol{r}}_{\mathrm{s}}(t) &= v_{s} \hat{\boldsymbol{z}},
\end{align}
where $\boldsymbol{r}_{\mathrm{tr}}$ and $\boldsymbol{r}_{\mathrm{s}}$ denote the tracer and swimmer positions, respectively, and the flow field $\boldsymbol{u}$ and the initial conditions are set by the geometry of interest (PBCs or spherical cavity), see below. For bulk (infinite and non-periodic) systems, we numerically approximate an infinite tracer trajectory by using a path length of $5 \times 10^{4}$, which we have previously shown to be sufficient to reach the bulk limit~\cite{degraaf16c}.

We start by considering the near field, for which the details of the force distribution matter. In order to accurately compare with simulations in this regime, we use two point forces placed a distance $l = 1$ apart, the so-called ``di-Stokeslet'' description, rather than a point stresslet. As a semi-empirical mathematical description of the ``smearing out'' of the force onto the LB lattice due to the force interpolation, we employ the regularized (non-singular) Stokeslet proposed by Cortez~\textit{et al.}~\cite{Cortez05}:
\begin{align}
\label{eq:Stokes-reg} \boldsymbol{S}_{\mathrm{reg}}(\boldsymbol{r};\varepsilon) &= \frac{ \left( r^{2} + 2\varepsilon^{2} \right) \mathbb{I} + \boldsymbol{r} \otimes \boldsymbol{r} }{ \left( r^{2} + 2\varepsilon^{2} \right)^{3/2} } ,
\end{align}
with $\mathbb{I}$ the 3D identity matrix, $\otimes$ the dyadic product, and the associated fluid velocity
\begin{align}
\boldsymbol{u}_{\mathrm{reg}}(\boldsymbol{r};\varepsilon) &= \frac{1}{8 \pi \mu}\mathbf{S}_{\mathrm{reg}}(\boldsymbol{r};\varepsilon) \boldsymbol{f} .
\end{align}
For finite $\varepsilon > 0$, this expression corresponds to a non-singular force density smeared out over a volume $\sim \varepsilon^{3}$, and in the limit $\varepsilon \rightarrow 0$ it reduces to the ordinary (singular) Stokeslet which is a fundamental solution to Eq.~\eqref{eq:Stokes}. 

In the far-field regime ($x_{i} \gg l$), where the details of the boundary conditions become important, we instead employ fluid velocities $\boldsymbol{u}(\boldsymbol{r})$ given by point-stresslet expressions either in PBCs through the Ewald sum derived in~\cite{degraaf16c} or in a spherical cavity, as shown in Appendix~\ref{sec:cavity}. For completeness, we also compare our results to the velocity field of the bulk stresslet expression given in Eq.~\eqref{eq:bulkStresslet}.

\section{\label{sec:results}Results}

In the following, we will assess the reliability of the LB simulations by comparing the tracer trajectories obtained in our numerical calculations with the corresponding theoretical estimates. First, we examine the effect of the short-range regularization imposed by the force-fluid coupling on tracer motion for small swimmer-tracer separations $x_{i}$. Next, we consider larger $x_{i}$, where the effect of the boundary conditions become significant. Finally, we study the effect of having a non-zero Reynolds number in the simulations and establish when and how the Stokes flow approximation starts to break down. 

\subsection{\label{sub:nearfield}Near-Field Flows and the Effect of Regularization}

\begin{figure}[!htb]
\centering 
  \includegraphics[scale=1.0]{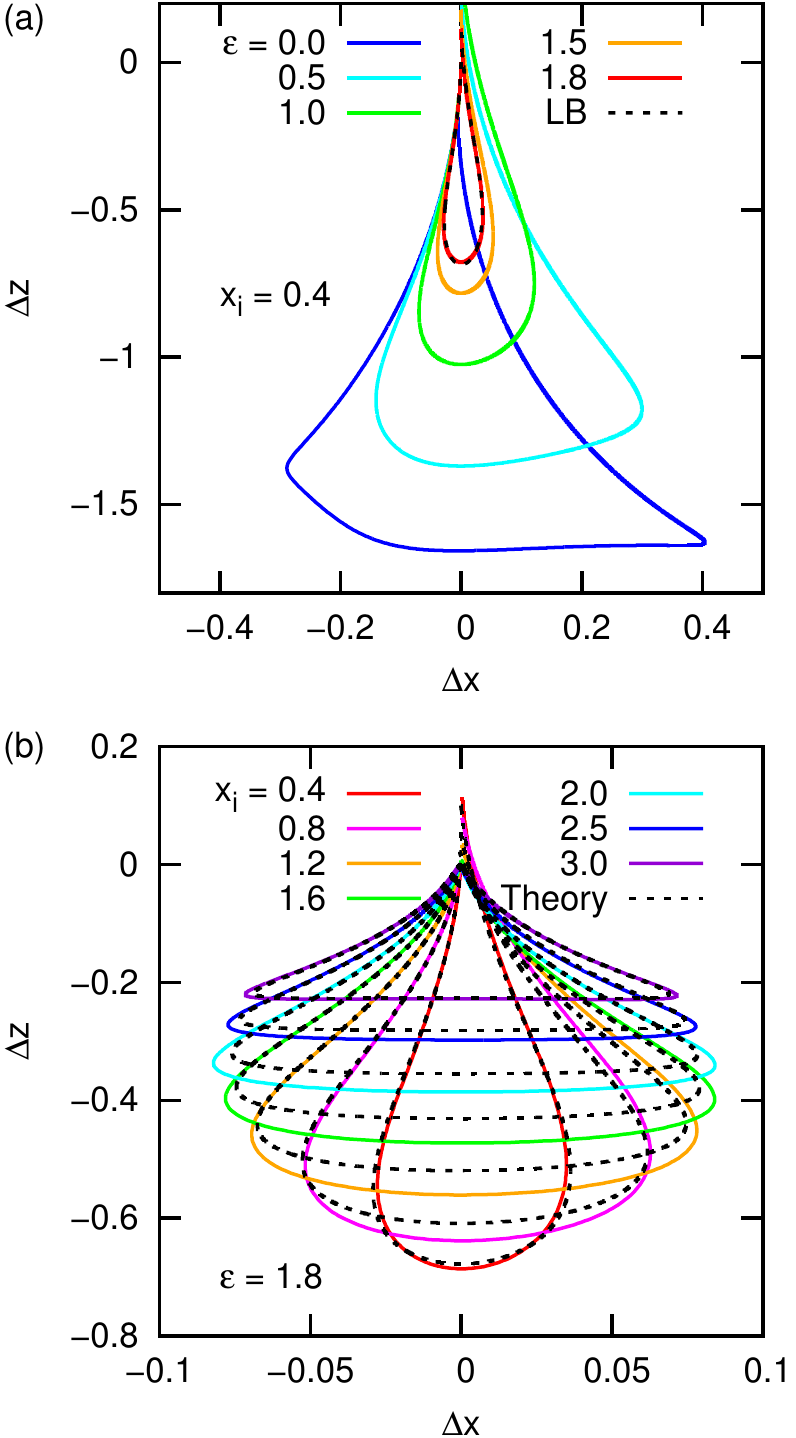} 
  \caption{\label{fig:near}Comparison between the near-field trajectories of regularized theory and LB simulations for a cubic system with PBCs and $L=100$. (a) Theoretical tracer trajectories for various values of the regularization parameter $\varepsilon$ (solid lines) and $x_{i} = 0.4$, together with the corresponding LB trajectory (dashed line). (b) Near-field advection of tracers by a puller swimmer for small values of $x_{i}$ obtained from LB simulations (solid lines) and from theoretical calculations using a pair of regularized Stokeslets (dashed lines) with $\varepsilon = 1.8$. For the theoretical curves, PBCs are included through a direct summation of images in spherical shells, which is computationally feasible for the very small swimmer-tracer separations ($x_{i} / L \leq 0.03$) considered here.}   
\end{figure} 

\begin{figure}[!h]
\centering 
  \includegraphics[scale=1.0]{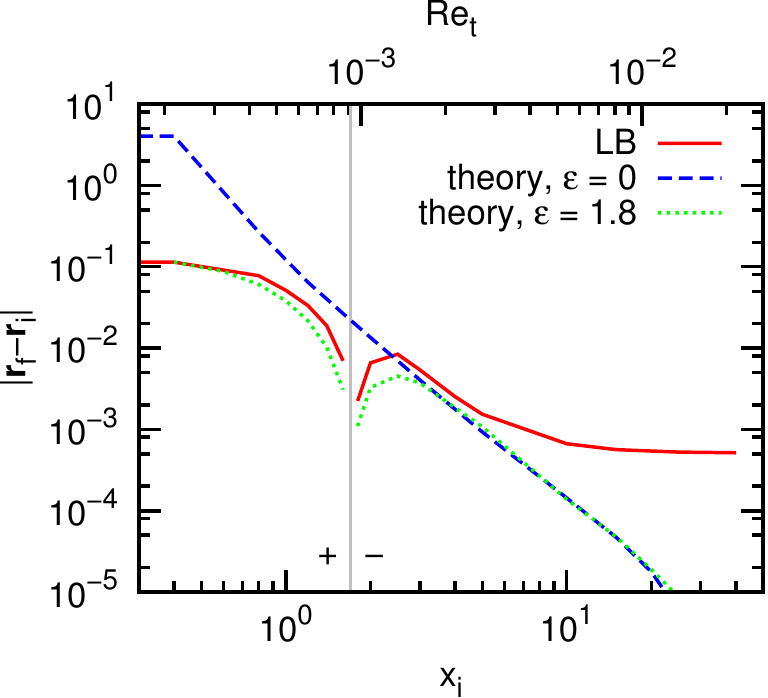} 
  \caption{\label{fig:entrain}The total net tracer displacement $\vert \boldsymbol{r}_{f} - \boldsymbol{r}_{i} \vert$ in a cubic system with PBCs ($L = 100$) as a function of the initial tracer position $x_{i}$. The red solid curve shows the LB result, the blue dashed curve indicates the non-regularized ($\varepsilon = 0$) Ewald sum result for a di-Stokeslet, and the green dotted curve shows the corresponding results using the regularization of Eq.~\eqref{eq:Stokes-reg} ($\varepsilon = 1.8$). For the regularized curve, we account for PBCs by a direct summation over spherical shells of the images of the swimmer. The cusps in the LB and regularized theory results are due to a sign inversion of the $z$-component of $\boldsymbol{r}_{f} - \boldsymbol{r}_{i}$, which gives the dominant contribution to the displacement. For the non-regularized result ($\varepsilon = 0$) this value is always negative, but for the LB and regularized curve, the $z$-component becomes positive close to the swimmer, as indicated by the gray vertical line and $+/-$ symbols. The top axis gives the Reynolds number corresponding to the initial tracer position $\mathrm{Re}_{t}$, as explained in Section~\ref{sub:Reynolds}.}
\end{figure}

The solid lines in Fig.~\ref{fig:near} show tracer trajectories for small swimmer-tracer separations ($x_{i} \leq 3$) in a system with PBCs and $L = 100$. The LB trajectories have distinctly different shapes compared to those obtained using non-regularized ($\varepsilon = 0$) di-Stokeslet theory. While the latter trajectories always have a concave component at the base, the LB curves are convex there. In addition, the net tracer displacement due to Darwin drift ~\cite{Darwin-1953, Pushkin-2013} ($\vert \boldsymbol{r}_{f} - \boldsymbol{r}_{i} \vert$ in Fig.~\ref{fig:entrain}) has the opposite sign (positive rather than negative) compared to the one predicted by non-regularized theory for small swimmer-tracer separations. Note that Darwin drift specifically refers to the permanent (net) displacement of a fluid parcel and hence tracer particle, due to the passing of the swimmer.

The origin of the discrepancies in the near-field flows between simulations and theory is the interpolation of the forces and fluid velocity between the off-lattice swimmers and tracers and the lattice fluid, see Fig.~\ref{fig:model}. This causes a ``smeared out'' flow field compared to that produced by point Stokeslets, which, while not being a realistic description of the flow-field from a real microswimmer, prevents divergences for short swimmer-tracer separations. To include this effective volume-force distribution into our theoretical curves, we employ the regularization proposed by Cortez~\textit{et al.}~\cite{Cortez05}, see Eq.~\eqref{eq:Stokes-reg}. In Fig.~\ref{fig:near}a, we fit the advection induced by a regularized di-Stokeslet using different values of $\varepsilon$ to the corresponding LB data at $x_{i} = 0.4$. We find excellent agreement for $\varepsilon = 1.8$,~\textit{i.e.}, a regularization length scale of about 2 lattice points. This is reasonable for a 3-point interpolation scheme, as the interpolation occurs over a region of size $2$, see Fig.~\ref{fig:model}. 

Note the extremely large effect the regularization has on the near-field advection, comparing the $\varepsilon = 0$ result to that of the LB ($\varepsilon = 1.8$) in Fig.~\ref{fig:near}, indicating that this regime is indeed not well-described by a non-regularized (extended or point) stresslet model. Figures~\ref{fig:near}b and~\ref{fig:entrain} show satisfactory correspondence between LB results and regularized theory over a wide range of separations, capturing the trend in the trajectory and change of sign in the $z$-component of the displacement well. This is remarkable, since the exact mathematical form of the regularization in the LB simulations is not known \textit{a priori}, and is not expected to be identical to the generic form of Eq.~\eqref{eq:Stokes-reg}. This observation is important to match theoretical predictions and simulations of the behavior of suspensions of microswimmers and tracers for this model. However, for the purposes of accurately modeling microorganisms, the region close to the swimmer will require near-field corrections that will likely dominate over this effect.

There is, however, a significant deviation between both sets of theoretical results and the LB simulations for intermediate values of $x_{i}$ in Fig.~\ref{fig:entrain}. We attribute this difference to the fact that the net displacement is very small compared to the extent of the trajectory and therefore much more sensitive to small changes in the latter. Therefore, it is also highly sensitive to numerical rounding errors and algorithmic details, such as the order of the interpolation and the use of floating-point arithmetic, of the GPU-based LB method.

\subsection{\label{sub:farfield}Far-field Flows and the Effect of Boundary Conditions}

\begin{figure*}[!htb]
\centering 
  \includegraphics[scale=1.0]{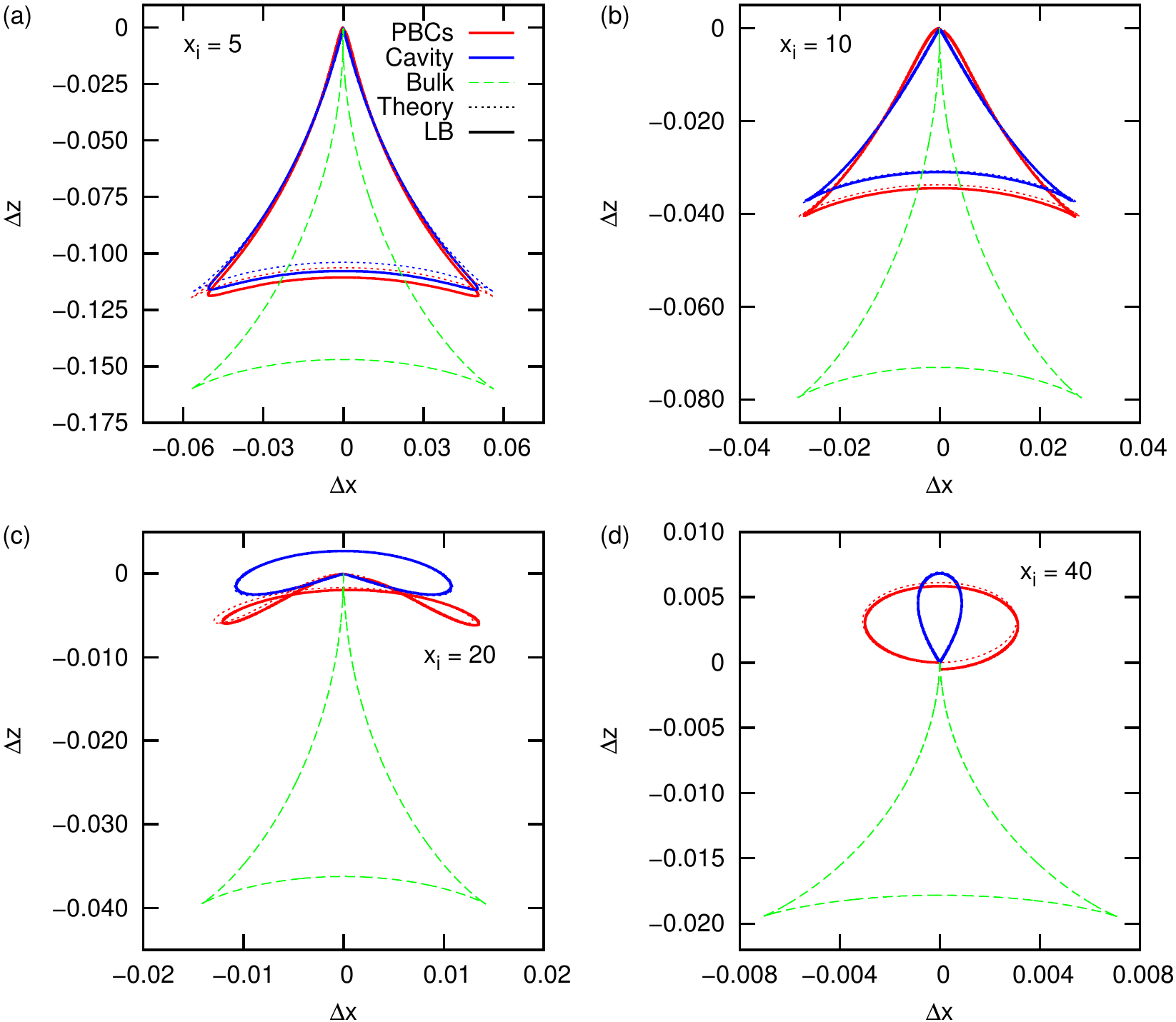}
  \caption{\label{fig:far}Advection of tracers by puller swimmers in three different geometries: A cubic box with PBCs (red), a no-slip spherical cavity (blue), and an infinite bulk fluid (green, dashed). For the cube and the cavity, the edge-length and diameter are the same ($L = 2a = 100$), and all other parameters are equal for the two systems. Solid lines indicate the LB result, while dotted curves show results of theoretical calculations based on solving the Stokes equation under the respective boundary conditions. The four panels show different initial tracer positions $x_{i}$, as indicated.}   
\end{figure*} 

We now turn to the far-field (large $x_{i}$) tracer trajectories, where we consider both a cubic system with PBCs and a finite spherical cavity with no-slip walls, see Fig.~\ref{fig:far}. Comparison between the tracer trajectories and the bulk results demonstrates that both sets of boundary conditions significantly affect the path followed by the tracer, even for swimmer-tracer separations that are small compared to the system dimensions ($x_{i}/ L$ and $x_{i} / a \ll 1$), in line with our previous observations~\cite{degraaf16c}. 

Furthermore, the trajectories obtained by LB match our theoretical calculations quantitatively, indicating that the regularization error present in the near-field trajectories is negligible for these separations. This also constitutes an independent verification of our Ewald-summed stresslet~\cite{degraaf16c} and confirms the expressions derived in Appendix~\ref{sec:cavity} for the spherical cavity. However, there are some subtleties to the LB trajectories when compared to the theoretical result, namely a slight skewness (particularly noticeable for $x_{i} = 20$ and PBCs) and a sizeable $\vert \boldsymbol{r}_{f} - \boldsymbol{r}_{i} \vert$ ($x_{i} = 40$ and PBCs), to which we will return in Section~\ref{sub:Reynolds}.

Finally, there are large similarities between the trajectories obtained from the PBC system and the one enclosed in a spherical cavity, while both of them differ much more from the trajectories in an infinite (bulk) system. Qualitatively, this can be understood as arising from a cutoff of the hydrodynamic modes for length-scales larger than the box dimensions in PBCs~\cite{Dunweg-1993, Ripoll-2005, Huang-2012}. It is nevertheless rather striking that the PBCs quantitatively very closely emulates the results found for a finite, spherically confined system, since the stresslet flow field in a system with PBCs is unaffected by the position of the swimmer, while the stresslet in a spherical cavity is strongly position-dependent due to the swimmer's proximity to the wall. Hence, the flow fields of a stresslet in both types of boundary condition are quantitatively different.

\begin{figure}[!h]
\centering 
  \includegraphics[scale=1.0]{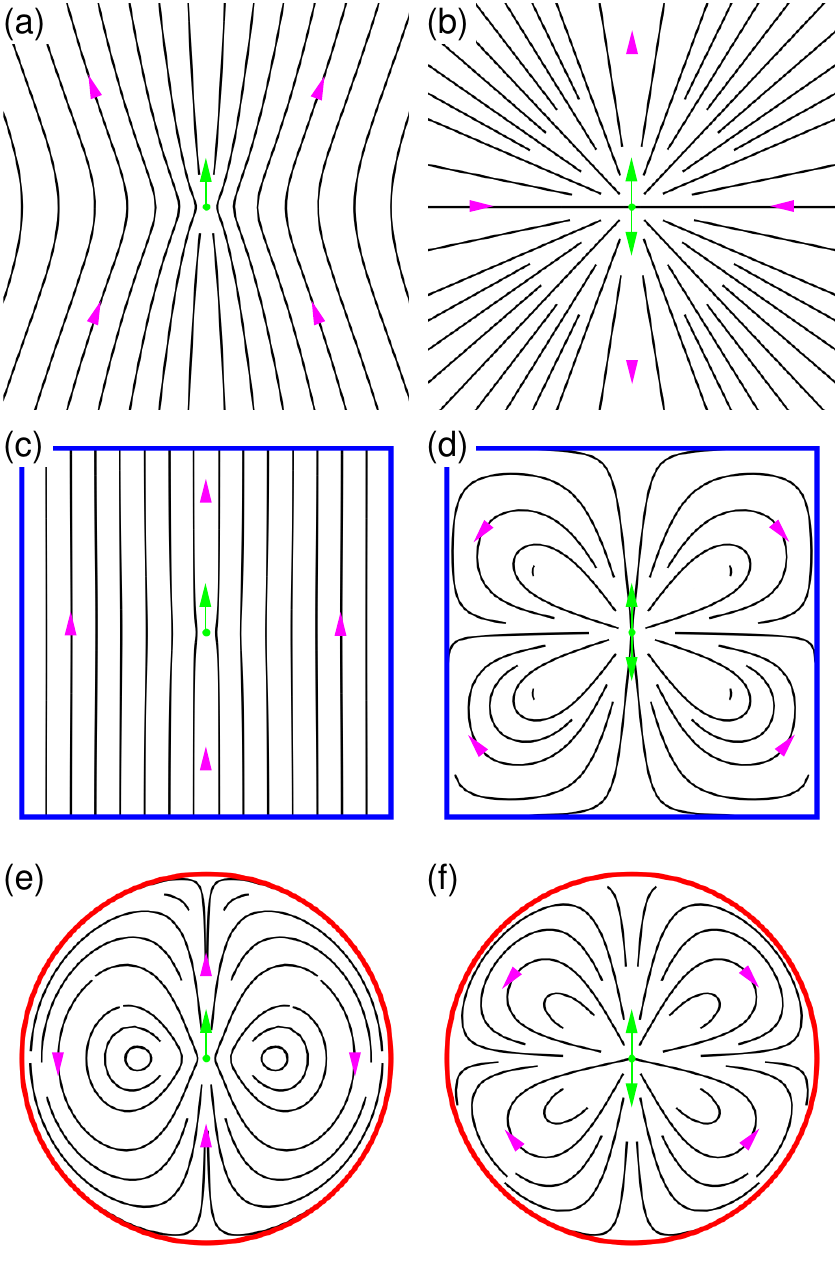} 
  \caption{\label{fig:topology}The shape of the flow field due to a Stokeslet (left) and stresslet (right) in different system geometries. The black curves give the flow lines, the magenta arrows the direction of the flow, the green dot and arrow the location and direction of the force (Stokeslet), and the green dot and double arrow the location of the stresslet. The considered systems are: (a,b) an infinite fluid volume (bulk); (c,d) a cube with PBCs, and (e,f) a spherical cavity with no-slip walls.}
\end{figure} 

At a more pictorial level, the similarity between the tracer trajectories in the two geometries can be understood by the topology of the flow field, see Fig.~\ref{fig:topology}. In an array of swimmers (corresponding to PBCs), the flow field of each swimmer either pushes or pulls on the flow coming from its neighbor. Due to incompressibility, this causes the flow to ``loop back'' on itself in much the same way as the flow loops back on itself when solid no-slip walls are used. That this is an effect of the symmetry of the dipolar flow field can be understood by making the corresponding comparison for a single Stokeslet. In PBCs, the Stokeslet flow field is ``unidirectional'' and neighboring force points do not cause the flow to loop back on itself, unlike the situation in a confined system. These far field ``loops'' in the flow field of the stresslet can thus be used to qualitatively explain the similarities between the two geometries. 

\subsection{\label{sub:Reynolds}Effects of Non-Zero Reynolds Numbers}

\begin{figure*}
\centering 
  \includegraphics[scale=1.0]{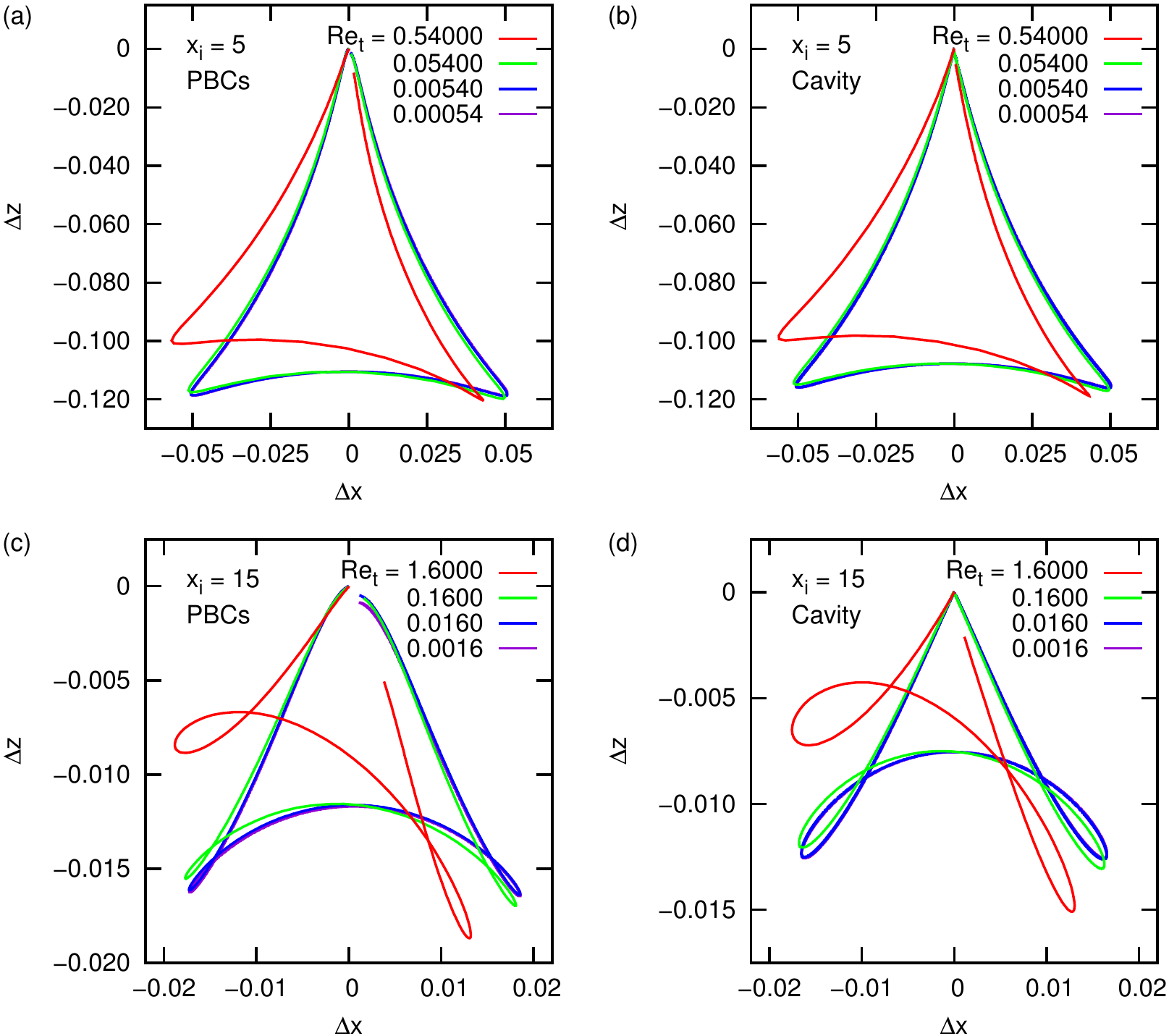} 
  \caption{\label{fig:retard} Retardation effects at non-zero Reynolds number as observed in LB simulations for a cubic system with PBCs (a,c) and a spherical cavity (b,d) with two different values of $x_{i}$, as indicated. The tracer Reynolds number is defined by Eq.~\eqref{eq:Re} with $\ell = x_{i}$.}   
\end{figure*}

Figure~\ref{fig:retard} shows LB tracer trajectories for the two different sets of boundary conditions for varying values of the swimming force and thus the swimming velocity $v_{s}$, in order to explore the effect of non-zero Reynolds numbers. To this end, we introduce the \textit{tracer} Reynolds number $\mathrm{Re}_{t}$, based on the swimmer-tracer separation $x_{i}$ rather than the swimmer length $l$ as the relevant length-scale $\ell$ in Eq.~\eqref{eq:Re}. By changing the swimming force, while keeping all other parameters fixed, we thus alter $\mathrm{Re}_{t}$ \textit{via} $v_{s}$~\footnote{Note that, for all curves in Fig.~\ref{fig:retard}, the \textit{swimmer} Reynolds number $\mathrm{Re}_{s} <  0.1$ for all values of $\mathrm{Re}_{t}$, which is made possible by the fact that $l \ll x_{i}$.}.

For $\mathrm{Re}_{t} \ll 1$ (the Stokes flow regime), all the tracer trajectories collapse onto each other. This is expected in the linear regime, because there reducing the driving force $f$ (or, equivalently, the stresslet strength $\kappa$) causes a corresponding reduction in swimming speed $v_{s}$ and thus leads only to a rescaling of the effective time unit of the problem. That is, a smaller tracer displacement (due to a reduced $\kappa$) acts over a longer time (due to a reduced $v_{s}$), and these effects exactly cancel each other out. 

As $\mathrm{Re}_{t}$ grows above $\approx 0.1$, the tracer trajectories start to become significantly skewed. As the time needed for fluid momentum to diffuse from the swimmer to the tracer becomes longer than the time needed for the swimmer to move a significant distance, an effective retardation of the swimmer-tracer interaction becomes visible. This retardation is present and similar in the system with PBCs and in the spherical cavity, as can be seen by comparing the left- and right-hand sides of Fig.~\ref{fig:retard}. The presence of momentum-absorbing walls does not appear to substantially impact the retardation experienced by the swimmer.

Finally, we should note that the retardation affects both the trajectory (advection) and the net displacement (Darwin drift) of the tracer. Retardation should be considered a separate effect, however, as both advection and Darwin drift are present for $\mathrm{Re} = 0$. Returning to Fig.~\ref{fig:entrain}, we can see that retardation more significantly impacts the net displacement than the shape of the trajectory itself, due to the greater sensitivity of this displacement to changes in the trajectory. 

\section{\label{sec:conclu}Conclusions}

In this paper, we have examined the trajectories of tracers that are advected in the flow field of a single, non-tumbling microswimmer in the absence of thermal fluctuations. We contrasted the results obtained using a lattice-Boltzmann method with force/counter-force swimmers against theoretical calculations in a cubic geometry with periodic boundary conditions and a spherical cavity with no-slip walls. We found that there are three main effects that need to be taken into account in comparing LB simulations to theoretical calculations: (i) the near-field flow, due to the lattice interpolation, (ii) the geometry of the fluid domain, and (iii) the finite (and relatively low) speed of momentum transport in the LB fluid.

The LB point-coupling algorithms of Refs.~\cite{nash08, degraaf16a} rely on an interpolation of the force and stress onto a lattice. This coupling leads to an inherent near-field regularization of the flow field compared to unregularized fundamental solutions to the Stokes equation. In studying the mean-squared displacement of tracers due to a bath of swimmers, one should take this significant near-field difference into account. We found that the near-field flow in the LB simulations can be well-approximated using a relatively simple theoretical regularization~\cite{Cortez05} with a regularization parameter obtained by matching to the LB tracer trajectories. 

In the far field, we observed a very significant influence of the type of boundary condition used. Interestingly, there is substantial similarity among trajectories in PBCs and in the confined system of the spherical cavity, while they differ much more from the trajectories in an infinite (bulk) system. This is due to the symmetry of the dipolar flow field, which causes a topological change in the stresslet flow from bulk to PBCs, namely the appearance of closed flow lines. Such loops are also present in the cavity, but this similarity between confinement and periodicity is not present for the Stokeslet. The result implies that, at least for the particular case of microswimmers, the use of PBCs to achieve a more bulk-like system actually gives rise to results that are more akin to those obtained in a confined geometry.

Furthermore, our results indicate that, for a reasonable approximation of non-inertial swimming, an effective Reynolds number less than $0.1$ is needed. This Reynolds number takes the speed of the swimmer and the size of the geometric feature of interest (in our case the swimmer-tracer separation, which may be comparable to the size of the simulation box). This upper bound on the Reynolds number for the accurate reproduction of the Stokes flow result is in accordance with the observations made previously~\cite{Cates-2004, nash08}, and agrees with similar limits obtained for non-swimming systems. In an LB simulation, the natural way of decreasing the Reynolds number is to simply lower the swimming velocity. However, for large-scale flows such as seen in studies of collective motion in bacterial suspensions, the length scales are large enough to potentially cause computational difficulties, as a smaller swimming velocity means that a larger number of time steps is needed to sample the same configuration space. 

Overall, our work demonstrates that the accurate simulation of hydrodynamic interactions between swimmers and tracers using LB and similar methods is a nontrivial matter. The specifics of the simulation domain and the choices for the swimmer speed and LB fluid parameters all have a very significant impact on the results, meaning that great care must be taken to recover the physics of the system of interest. 
\section*{\label{sec:acknowledgements}Acknowledgements}

We would like to thank Alexander Morozov and Rupert Nash for helpful discussions. JdG thanks the ``Deutsche Forschungsgemeinschaft'' (DFG)  for funding  through  the  SPP 1726  ``Microswimmers: from  single  particle  motion  to  collective behavior'' (HO1108/24-1) and gratefully acknowledges funding by a Marie Sk{\l}odowska-Curie Intra European Fellowship (G.A. No. 654916) within Horizon 2020. JS is financed by a Project grant from the Swedish Research Council (2015-05449).

\bibliographystyle{aip}
\bibliography{LB}

\appendix
\section{\label{sec:cavity}Flow field of Stokeslets and Stresslets in a Spherical Cavity}

In this section, we compute the velocity field $\boldsymbol{u}$ induced by a point force monopole (Stokeslet) and symmetric force dipole (stresslet) on the fluid confined in a spherical cavity with no-slip walls in the laminar-flow (low-$\mathrm{Re}$) regime described by Eq.~\eqref{eq:Stokes}. To compute this Stokeslet, we assume that a point force $\boldsymbol{f}$ is directed along the symmetry axis $\boldsymbol{\hat{z}}$ of the system and employ spherical polar coordinates (SPCs) with polar angle $\theta$. Due to the axisymmetric nature of the flow we are interested in, all relations will be independent of the azimuthal angle $\phi$. We use the stream function approach in our calculations, which allows us to solve for the stream lines --- contours of the stream function --- of the Stokes equation~\eqref{eq:Stokes}. In the axisymmetric case, closed analytical expressions can be derived~\cite{Collins58}. The computation for an arbitrarily directed force is much more involved, and is furthermore not relevant to the comparisons in this manuscript, and is therefore not considered here. Once the correct Stokeslet expression has been established, the stresslet is derived from it by taking the directional derivative with respect to the location of the applied force. 

\subsection{\label{subsec:bulk}Stream Function of a Stokeslet in an Infinite Fluid}

We begin by considering the stream function $\psi_{0}(r,\theta)$ to the Stokes equation~\eqref{eq:Stokes} in SPCs for an infinite fluid domain. The stream function can be shown to fulfil the differential equation~\cite{lamb45, Acheson90}
\begin{align}
\label{eq:streamdiff} \left[ \frac{\partial^{2}}{\partial r^{2}} + \frac{\sin \theta}{r^{2}} \frac{\partial}{\partial \theta} \left( \frac{1}{\sin \theta} \frac{\partial}{\partial \theta} \right) \right]^{2} \psi_{0}(r,\theta) &\equiv \Lambda^{4} \psi_{0}(r,\theta) = 0,
\end{align} 
with the differential operator 
\begin{align}
\label{eq:Lambda2} \Lambda^{2} &= \frac{\partial^{2}}{\partial r^{2}} + \frac{\sin \theta}{r^{2}} \frac{\partial}{\partial \theta} \left( \frac{1}{\sin \theta} \frac{\partial}{\partial \theta} \right) .
\end{align}
A stream function that satisfies Eq.~\eqref{eq:streamdiff} allows us to write
\begin{align}
\label{eq:strur} u_{r}(r,\theta)      &=  \frac{1}{r^{2} \sin \theta} \frac{\partial}{\partial \theta} \psi_{0}(r,\theta) ; \\
\label{eq:strut} u_{\theta}(r,\theta) &= -\frac{1}{r \sin \theta} \frac{\partial}{\partial r} \psi_{0}(r,\theta) ,
\end{align}
for the radial and tangential components of the fluid velocity, respectively.

Straightforward algebra, see Refs.~\cite{lamb45, Acheson90}, reveals that the stream function due to a point force $\boldsymbol{f} = f \boldsymbol{\hat{z}}$ applied in the origin ($\boldsymbol{r} = \boldsymbol{0}$) is given by
\begin{align}
\label{eq:psiorev} \psi_{0}(r,\theta) &= \frac{f}{8 \pi \mu} r \sin^{2} \theta .
\end{align}
Using Eqs.~\eqref{eq:strur} and~\eqref{eq:strut} we obtain for the flow field
\begin{align}
\label{eq:strureval} u_{r}(r,\theta)      &=  \frac{f}{8\pi\mu} \frac{2 \cos \theta}{r} ; \\
\label{eq:struteval} u_{\theta}(r,\theta) &= -\frac{f}{8\pi\mu} \frac{\sin \theta}{r} ,
\end{align}
which is simply the bulk Stokeslet in SPCs.

We now let the force $f \boldsymbol{\hat{z}}$ be applied at the point $z_{s} \boldsymbol{\hat{z}}$ instead of at the origin. Application of Pythagoras' theorem then leads to the following expression for the associated stream function, with an explicit parametric dependence on $z_{s}$:
\begin{align}
\label{eq:psi} \psi_{0}(r,\theta;z_{s}) &= \frac{f}{8 \pi \mu} \frac{r^{2} \sin^{2} \theta}{\sqrt{r^{2} - 2 r z_{s} \cos \theta + z_{s}^{2}}},
\end{align}
where $r$ and $\theta$ still specify the position of the point of interest in the fluid with respect to the origin. 

\subsection{\label{subsec:cavityStokeslet}Stream Function of a Stokeslet in a Spherical Cavity}

Next, we convert the stream function $\psi_{0}(r,\theta;z_{s})$ for the bulk axisymmetric system into a stream function $\psi_{c}(r,\theta;z_{s})$ for a spherical cavity of radius $a$ with no-slip walls, centered on the origin --- the subscript ``$c$'' indicates the cavity geometry. We apply the result by Collins~\cite{Collins58} to write
\begin{align}
\nonumber \psi_{c}(r,\theta;z_{s}) &= \psi_{0}(r,\theta;z_{s}) + \frac{ r \left( r^{2} - 3a^{2} \right) }{ 2a^{3} } \psi_{0}\left( \frac{a^{2}}{r} , \theta; z_{s} \right) \\
\nonumber                    &\quad  + \frac{ r^{2} \left( r^{2} - a^{2} \right) }{ a^{3} } \frac{\partial}{\partial r} \psi_{0}\left( \frac{a^{2}}{r} , \theta; z_{s}\right) \\
\label{eq:psisphdef}            &\quad  - \frac{ r^{2} \left( r^{2} - a^{2} \right)^{2} }{ 4 a^{5} } \Lambda^{2} \left[ r \psi_{0}\left( \frac{a^{2}}{r} , \theta; z_{s} \right) \right] .
\end{align}
Plugging in Eq.~\eqref{eq:psi} and evaluating the various expressions in Eq.~\eqref{eq:psisphdef} yields
\begin{align}
\label{eq:psisph} \psi_{c}(r,\theta; z_{s}) &= \frac{ r \left( A_{2} + A_{3} + A_{4} + A_{5} \right) \sin^{2} \theta  }{ 2 A_{0} A_{1} (a^{4} + r^{2} z_{s}^{2} - 2 a^{2} r z_{s} \cos \theta ) }; \\
\label{eq:aux0} A_{0}(r,\theta; z_{s}) &= \sqrt{r^{2} + z_{s}^{2} - 2 r z_{s} \cos \theta } ; \\
\label{eq:aux1} A_{1}(r,\theta; z_{s}) &= \sqrt{ \frac{a^{4}}{r^{2}} + z_{s}^{2} - \frac{2 a^{2} z_{s} \cos \theta }{r} } ; \\
\label{eq:aux2} A_{2}(r,\theta; z_{s}) &= -3 a^{5} A_{0} + 2 a^{4} A_{1} r ; \\
\label{eq:aux3} A_{3}(r,\theta; z_{s}) &= -3 a A_{0} r^{2} z_{s}^{2} + 2 A_{1} r^{3} z_{s}^{2} ; \\
\label{eq:aux4} A_{4}(r,\theta; z_{s}) &= a^{3} A_{0} (r^{2} + z_{s}^{2}); \\ 
\label{eq:aux5} A_{5}(r,\theta; z_{s}) &= 4 a^{2} r \left( a A_{0} - A_{1} r \right) z_{s} \cos \theta,
\end{align}
where the $A_{i}$ are auxiliary functions and we have dropped the functional dependencies of the $A_{i}$ on the right-hand side to ease the notation.

\begin{figure}[!htb]
\centering 
  \includegraphics[scale=1.0]{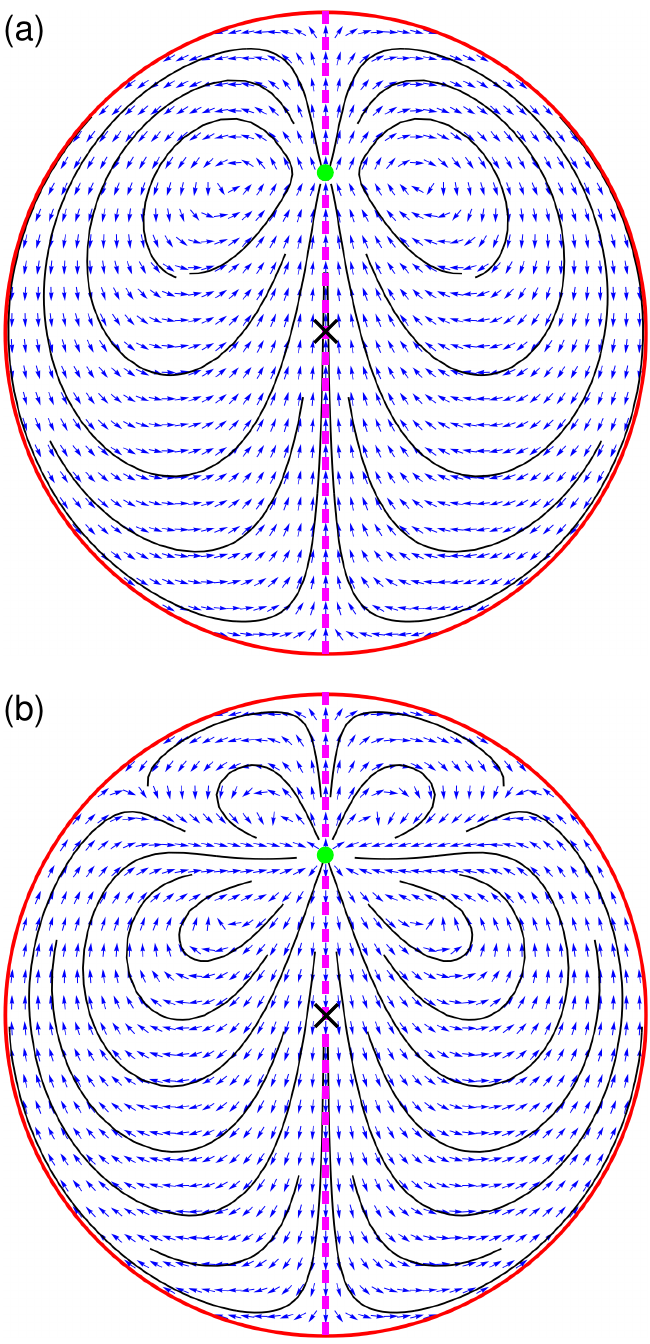} 
  \caption{\label{fig:cavity} Representations of the flow field induced in a spherical cavity with no-slip boundary conditions by (a) an upwards-pointing Stokeslet and (b) a puller stresslet pointing along the rotational symmetry axis (dashed magenta line; $\boldsymbol{\hat{z}}$). The position of the singularity is indicated using a green dot ($z = z_{s}$), the no slip-walls are represented by a red line, the flow field by (normalized) blue arrows, and the stream lines by black curves. The $\boldsymbol{\times}$ symbol shows the origin.}   
\end{figure}

The stream lines generated by Eq.~\eqref{eq:psisph} are shown in Fig.~\ref{fig:cavity}a. Using the stream function of Eq.~\eqref{eq:psisph} we can now write the radial and tangential components of the velocity field inside the sphere due to a point force applied at $-a < z_{s} < a$ as (\textit{c.f.}, Eqs.~\eqref{eq:strur}-\eqref{eq:strut})
\begin{align}
\label{eq:strursph} u_{f,r}(r,\theta; z_{s})      &=  \frac{1}{r^{2} \sin \theta} \frac{\partial}{\partial \theta} \psi_{c}(r,\theta; z_{s}) ; \\
\label{eq:strutsph} u_{f,\theta}(r,\theta; z_{s}) &= -\frac{1}{r \sin \theta} \frac{\partial}{\partial r} \psi_{c}(r,\theta; z_{s}),
\end{align}
where the subscript ``$f$'' indicates that the velocity derives from a force. This velocity field is shown in Fig.~\ref{fig:cavity}a using blue arrows.

\subsection{\label{subsec:cavityStresslet}Stream Function of a Stresslet in a Spherical Cavity}

The flow field of an extended stresslet (di-Stokeslet), composed of two inverted Stokeslets of equal magnitude located at $z = z_{s} \pm l/2$, is simply the sum of two expressions similar to those in Eqs.~\eqref{eq:strursph} and~\eqref{eq:strutsph}. Assuming a constant dipole strength $\kappa \equiv f l$, and taking the limit of $l \rightarrow 0$, the flow field $u_{s}$ due to the point stresslet is obtained by the directional derivative with respect to $z_{s}$. This yields
\begin{align}
\label{eq:strursphstr} u_{s,r}(r,\theta; z_{s})      &=   \frac{l}{r^{2} \sin \theta} \frac{\partial}{\partial \theta}  \psi_{c}'(r,\theta; z_{s}) ; \\
\label{eq:strutsphstr} u_{s,\theta}(r,\theta; z_{s}) &= - \frac{l}{r \sin \theta} \frac{\partial}{\partial r} \psi_{c}'(r,\theta; z_{s}) ; \\
\label{eq:derpsi}         \psi_{s}'(r,\theta; z_{s}) &\equiv \frac{\partial}{\partial z_{s}} \psi_{c}(r,\theta; z_{s}) ,
\end{align}
where the subscript ``$s$'' indicates that the velocity derives from a stresslet. By construction, positive values of $\kappa$ correspond to pusher swimmers and negative values to puller swimmers. The flow field and stream lines generated by a puller that is off-center with respect to the cavity are shown in Fig.~\ref{fig:cavity}b. The full expressions for $u_{s,r}(r,\theta; z_{s})$ and $u_{s,\theta}(r,\theta; z_{s})$ are not provided here, as they are very unwieldy.

\end{document}